\documentclass[aps,prl,twocolumn,superscriptaddress,showpacs,floatfix,10pt]{revtex4-1}

\usepackage[utf8]{inputenc}
\usepackage{microtype}

\usepackage{amsmath}
\usepackage{amssymb}
\usepackage{bm}

\renewcommand{\vec}{\bm}
\newcommand{\dif}{\mathrm{d}}
\newcommand{\me}{\mathrm{e}}
\newcommand{\mi}{\mathrm{i}}

\usepackage{graphicx}

\usepackage[pdfstartview=FitH,colorlinks=true,linkcolor=blue,citecolor=blue,urlcolor=blue]{hyperref}

\begin{document}

\title{Tricriticalities and Quantum Phases in Spin-Orbit-Coupled Spin-\texorpdfstring{$1$}{1} Bose Gases}

\author{Giovanni I. Martone}
\email{Giovanni.Martone@ba.infn.it}
\affiliation{Dipartimento di Fisica and MECENAS, Universit\`{a} di Bari, I-70126 Bari, Italy}
\affiliation{Istituto Nazionale di Fisica Nucleare (INFN), Sezione di Bari, I-70126 Bari, Italy}

\author{Francesco V. Pepe}
\affiliation{Museo Storico della Fisica e Centro Studi e Ricerche ``Enrico Fermi,'' I-00184 Roma, Italy}
\affiliation{Istituto Nazionale di Fisica Nucleare (INFN), Sezione di Bari, I-70126 Bari, Italy}

\author{Paolo Facchi}
\affiliation{Dipartimento di Fisica and MECENAS, Universit\`{a} di Bari, I-70126 Bari, Italy}
\affiliation{Istituto Nazionale di Fisica Nucleare (INFN), Sezione di Bari, I-70126 Bari, Italy}

\author{Saverio Pascazio}
\affiliation{Dipartimento di Fisica and MECENAS, Universit\`{a} di Bari, I-70126 Bari, Italy}
\affiliation{Istituto Nazionale di Fisica Nucleare (INFN), Sezione di Bari, I-70126 Bari, Italy}

\author{Sandro Stringari}
\affiliation{INO-CNR BEC Center and Dipartimento di Fisica, Universit\`{a} di Trento, I-38123 Povo, Italy}

\date{\today}

\begin{abstract}
We study the zero-temperature phase diagram of a spin-orbit-coupled Bose-Einstein condensate of spin $1$,
with equally weighted Rashba and Dresselhaus couplings. Depending on the antiferromagnetic or ferromagnetic
nature of the interactions, we find three kinds of striped phases with qualitatively different behaviors in the
modulations of the density profiles. Phase transitions to the zero-momentum and the plane-wave phases can
be induced in experiments by independently varying the Raman coupling strength and the quadratic Zeeman field.
The properties of these transitions are investigated in detail, and the emergence of tricritical points,
which are the direct consequence of the spin-dependent interactions, is explicitly discussed.
\end{abstract}


\maketitle

\textit{Introduction ---} Ultracold atoms with spin-orbit (SO) coupling represent a very active field
of research within the physics of quantum gases. They are characterized by highly nontrivial phase
structures, arising when their peculiar single-particle properties are combined with the effects of the
interparticle interactions. In the case of Bose gases, this results, e.g., in the appearance of
spin-polarized phases as well as of spatially modulated (striped) configurations with supersolid-like
features~\cite{Boninsegni2012} (see the recent reviews~\cite{Dalibard2011review,Galitski2013review,
Zhou2013review,Goldman2014review,Zhai2015review,Li2015review,Zhang2016review} and references therein).

Since the first achievement of a SO-coupled Bose-Einstein condensate (BEC) by the NIST group~\cite{Lin2011},
most studies have focused on spin-$1/2$ systems. Recent experiments in a gas of $^{87}$Rb atoms~\cite{Campbell2016,Luo2016}
have succeeded in implementing spin-orbit coupling on spin-$1$ BECs, thus paving the way towards the exploration of the
interplay between the spinor character and the SO coupling in determining the phase diagram of the system.
Some properties of SO-coupled BECs with higher spin have been investigated in~\cite{Wang2010,Xu2012,Ruokokoski2012,Lan2014,
Song2014,Gautam2014,Gautam2015a,Natu2015,Gautam2015b,Pixley2016,Chen2016a,Chen2016b,Hurst2016}.

The purpose of this Letter is to analyze the ground state of a SO-coupled BEC of spin $1$ in uniform matter
and to point out the emergence of quantum tricriticalities, which are approachable in a highly controllable way
by varying the Raman coupling and the quadratic Zeeman field. Tricriticality is an ubiquitous phenomenon
in multicomponent systems~\cite{Widom1996}, like superconductors~\cite{Kleinert1989,Green2005,Giovannetti2011},
antiferromagnets~\cite{Kato2015} and QCD configurations~\cite{Stephanov1998}. Experimental evidence of quantum
tricriticality has been recently reported in ferromagnetic superconductors~\cite{Tokunaga2015}. Controllability
is crucial to reveal experimentally the predicted unconventional critical behavior of systems close to tricriticality,
including the temperature dependence of the observables and their critical exponents~\cite{Mercaldo2011,Kato2014,Kato2015},
which can lead to the identification of novel universality classes~\cite{Kato2014}.

\textit{Single-particle physics ---} Let us consider a spin-$1$ BEC of $N$ particles in a volume $V$, characterized
by equally weighted Rashba~\cite{Bychkov1984} and Dresselhaus~\cite{Dresselhaus1955} SO couplings.
The single-particle Hamiltonian reads (we set $\hbar = m = 1$)
\begin{equation}
h_0 = \frac{1}{2} \left(p_x - k_0 F_z\right)^2 + \frac{p_\perp^2}{2}
+ \frac{\Omega}{2} \, F_x + \frac{\delta}{2} \, F_z +
\frac{\varepsilon}{2} \, F_z^2 \, ,
\label{eq:h0}
\end{equation}
where $p_{\perp}^2=p_y^2+p_z^2$ and $\vec{F}=\left(F_x,F_y,F_z\right)$ is the spin-$1$ operator. Hamiltonian~\eqref{eq:h0}
describes an atom coupled to two Raman laser beams (see, e.g., the experiment~\cite{Campbell2016}), within the rotating-wave
approximation and in a frame connected to the laboratory frame by a unitary transformation $U$, depending on space, time and
$F_z$~\cite{Note1}, and consisting of a proper generalization of the transformation employed in spin-$1/2$ gases~\cite{Martone2012}.
This transformation, with an additional renormalization of the quadratic Zeeman term, yields the time-independent and translationally
invariant Hamiltonian~\eqref{eq:h0}~\cite{Campbell2016,Campbell2015thesis}. The Raman coupling, quantified by $\Omega$, induces
transitions between the three spin components of the spin-$1$ BEC, providing at the same time a momentum transfer $k_0$ along $x$.
The spinor nature of the system enables one to introduce effective linear and quadratic Zeeman terms, with strength $\delta$
and $\varepsilon$, respectively, which can be tuned independently from each other by varying the frequencies of the Raman lasers.
Henceforth, we will focus on the $\delta=0$ case and study the interplay of Raman and quadratic Zeeman terms with interactions.

The single-particle spectrum of~\eqref{eq:h0} has been analyzed by Lan and \"{O}hberg in~\cite{Lan2014} and is characterized
by three branches. At small $\Omega$ and for large positive values of $\varepsilon$, the lowest branch has a single minimum at $p_x=0$.
At intermediate values of $\varepsilon$ two local minima appear at $p_x = \pm k_1$, with $0 < k_1 < k_0$, which become global degenerate
minima when $\varepsilon$ is further decreased. The minimum at $p_x = 0$ eventually disappears for large negative $\varepsilon$.
By increasing $\Omega$, the two curves delimiting the three-minima regime in the $\Omega$-$\varepsilon$ plane and the line where
the three minima are degenerate eventually merge in a critical point $C_S$, beyond which a direct transition from the single-minimum
to the double-minimum structure occurs.

\textit{Interacting gas ---} Two-body interparticle interactions introduce important novel features in the phase diagram. To investigate
the properties of the interacting gas, we resort to the Gross-Pitaevskii (GP) mean-field approach. The ground-state properties of the
system will be analyzed by minimizing the energy functional
\begin{equation}
E\left[\Psi\right] = \int_V \dif\vec{r} \left\{ \Psi^\dagger h_0 \Psi
+ \frac{g_0}{2} \left(\Psi^\dagger \Psi\right)^2
+ \frac{g_2}{2} \left(\Psi^\dagger \vec{F} \Psi\right)^2\right\} \, ,
\label{eq:E_Psi}
\end{equation}
where $\Psi$ is the three-component condensate wave function, with each component corresponding to the eigenvalues
of $F_z=\mathrm{diag}(+1,0,-1)$. The density $n(\vec{r}) = \Psi^\dagger(\vec{r}) \Psi(\vec{r})$ of the gas obeys the
normalization constraint $\int_V \dif\vec{r} \, n(\vec{r}) = N$. The coupling constants $g_0 = 4\pi(a_0+2a_2)/3$ and
$g_2 = 4\pi(a_2-a_0)/3$ are related to the $s$-wave scattering lengths in the spin-$0$ ($a_0$) and the spin-$2$ ($a_2$)
channels~\cite{Ho1998,Ohmi1998,StamperKurn2013}. The form of the two-body interaction in~\eqref{eq:E_Psi} is
characteristic of spin-$1$ BECs and reflects their rotational invariance. In the following, we shall focus on the ratios $g_2/g_0$
typical of ${}^{23}$Na ($g_2 > 0$) and ${}^7$Li ($g_2 <0$)~\cite{StamperKurn2013}.

The stationary condensate wave functions can be found by solving the GP equation
$\delta E / \delta \Psi^\dagger = \mu \Psi$, with $\mu$ the chemical potential. Since we are interested in the ground state,
we will consider solutions represented by a superposition of plane waves of the form
\begin{equation}
\Psi\left(\vec{r}\right) = \sqrt{\bar{n}} \,
\sum_{l \in \mathbb{Z}} C_l \, \Phi_l \, \me^{\mi l k_1 x} \, ,
\label{eq:ansatz}
\end{equation}
where $\bar{n} = N/V$ is the average density, $k_1$ the condensation momentum, $\Phi_l$ are real three-component
normalized spinors, and $C_l$ are complex coefficients obeying the normalization condition $\sum_{l \in \mathbb{Z}}
\left|C_l\right|^2 = 1$. The reality of $\Phi_l$ is related to the fact that the expectation value $\langle F_y\rangle$
in a stationary state of~\eqref{eq:E_Psi} vanishes at $\Omega \neq 0$~\cite{Note2}. To motivate the form~\eqref{eq:ansatz},
first notice that it reproduces the ground-state wave function for a noninteracting gas, obtained by solving the linear
Schr\"{o}dinger equation for the single-particle Hamiltonian~\eqref{eq:h0}. In that case, only the terms with $|l| \leq 1$
can be nonvanishing (see above); since the equation is linear, in the regimes where degenerate minima are present,
the relative values of the corresponding coefficients $C_l$ are arbitrary.

As we shall see, interactions lift this ground-state degeneracy, just as in spin-$1/2$ SO-coupled systems~\cite{Ho2011,Li2012PRL}.
We have studied the ground state of the system investigating stationary solutions of the GP equation in the
form~\eqref{eq:ansatz} as functions of the relevant parameters $k_0$, $\Omega$, $\varepsilon$, $\bar{n}$ and $g_2/g_0$.
We find quantum phases characterized by condensation both in a single momentum state (plane-wave phases)
and in a superposition of states with momenta $l k_1$ (striped phases), whose coefficients satisfy $|C_{-l}| = |C_l|$.
The contribution of the terms with $|l|>1$ in Eq.~\eqref{eq:ansatz}, appearing as a consequence of the nonlinearity
of the GP theory~\cite{Li2013}, is negligible for low-density systems, and becomes more and more relevant as $\bar{n}$
increases. Our results for the striped phases agree with~\cite{Lan2014}.

The global spin configuration of a spin-$1$ system can generally be described by means of two tensors,
$\langle \vec{F} \rangle$ and $\langle\mathbb{N}_{ij}\rangle$, related to two competing kinds of spin
order: ferromagnetic and nematic order, respectively~\cite{Mueller2004,StamperKurn2013}. The expectation
value $\langle \vec{F} \rangle$ of the spin operator is the magnetization vector, while the expectation value
$\langle\mathbb{N}_{ij}\rangle$ of the spin-quadrupole tensor
$\mathbb{N}_{ij} = \delta_{ij}\vec{F}^2/3 - (F_i F_j + F_j F_i)/2$, with $i,j=x,y,z$, is the nematicity tensor.
The eigenvectors belonging to positive eigenvalues of $\langle\mathbb{N}_{ij}\rangle$ identify the nematic directors.
One can also define magnetization and nematicity densities $\mathcal{F}_i(\vec{r}) = \Psi^\dagger(\vec{r}) F_i \Psi(\vec{r})$
and $\mathcal{N}_{ij}(\vec{r}) = \Psi^\dagger(\vec{r}) \mathbb{N}_{ij} \Psi(\vec{r})$. The sign of the spin-dependent
coupling $g_2$ plays a crucial role in establishing which kind of order characterizes the ground state of the
system~\cite{Ho1998,Ohmi1998,StamperKurn2013}.

\begin{figure}
\includegraphics{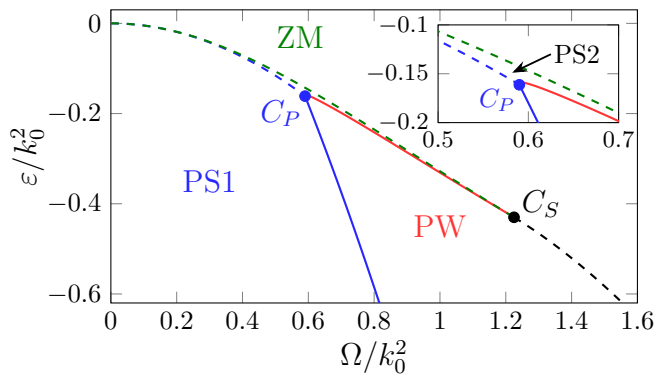}
\caption{Phase diagram for \emph{antiferromagnetic interactions} for $g_0\bar{n}/k_0^2 = 0.4$ and $g_2\bar{n}/k_0^2 = 0.012$.
The ratio $g_2/g_0 = 0.03$ is typical of ${}^{23}$Na. Continuous (dashed) lines represent first-(second-)order transitions.
The blue lines define the region of the first polar striped (PS1) phase. The region of the second polar striped (PS2) phase is delimited
by the dashed green and blue lines and by the continuous red one. The black line represents the transition between the zero-momentum
(ZM) and plane-wave (PW) phases. The inset displays a zoom close to the tricritical point $C_P$.}
\label{fig:polar_phase_diag}
\end{figure}

\textit{Antiferromagnetic interactions ---} Antiferromagnetic (or polar) spin-dependent interactions ($g_2>0$) tend to minimize the spin
term of Eq.~\eqref{eq:E_Psi}. Without SO coupling, if $\varepsilon > 0$, the ground state of the system is represented by the
longitudinal polar phase with wave function $\Psi = \sqrt{\bar{n}} \, (0 , 1 , 0)^T$, while for $\varepsilon < 0$ one has the transverse polar phase
$\Psi = \sqrt{\bar{n}/2}\,(1 , 0 ,\me^{2 \mi \phi})^T$, with $\phi$ an arbitrary phase factor~\cite{Ho1998,Ohmi1998,StamperKurn2013}.
The two phases are separated by a first-order transition at $\varepsilon = 0$. They both have $\langle \vec{F} \rangle=0$ and a
single nematic director, which is parallel to the $z$ axis for the longitudinal polar phase, and lies in the $x$-$y$ plane for the transverse
polar state, with its direction fixed by $\phi$.

The phase diagram in the presence of SO coupling is shown in Fig.~\ref{fig:polar_phase_diag}, as a function of the Raman coupling $\Omega$
and the quadratic Zeeman field $\varepsilon$. The upper part of the diagram is dominated by a zero-momentum (ZM) phase, with $C_0=1$.
Typical features of this phase are uniform density and vanishing longitudinal magnetization $\langle F_z \rangle$. At small $\Omega$ and
positive $\varepsilon$, the ground state approaches the above longitudinal polar state.

The lower part of the diagram of Fig.~\ref{fig:polar_phase_diag} corresponds to the region where the lowest branch of the single-particle spectrum
has two degenerate minima. The competition between density and spin-density interactions in~\eqref{eq:E_Psi} breaks this degeneracy. At low $\Omega$,
a polar striped (PS1) phase exists, characterized by spatial modulations and vanishing longitudinal magnetization $\langle F_z \rangle=0$. In this phase, only
harmonic terms with odd $l$ are present in the wave function~\eqref{eq:ansatz}, and the two states with momenta $\pm k_1$ are predominantly populated.
Stripe modulations with wavelength $\pi/k_1$ appear in the total density $n$, in the magnetization densities $\mathcal{F}_x$, $\mathcal{F}_y$,
and in the nematicity densities $\mathcal{N}_{xx}$, $\mathcal{N}_{yy}$, $\mathcal{N}_{xy}$, and $\mathcal{N}_{zz}$. Notice that, in the $\Omega \to 0$
limit, the polar striped phase approaches the transverse polar phase discussed above. The PS1 phase keeps a dominantly transverse polar character also at finite
$\Omega$, since the nematicity densities $\mathcal{N}_{xx}$, $\mathcal{N}_{yy}$, and $\mathcal{N}_{xy}$ are the observables that oscillate with the largest
contrast (see Supplemental Material~\cite{Supplemental}). However, the oscillations in $\mathcal{F}_x$ and $\mathcal{F}_y$ can augment significantly at large
$\Omega$. All these quantities are not invariant under $U^{-1}$, and exhibit a different space and time dependence in the laboratory frame~\cite{Martone2012}.
The total density $n$ and the nematicity density $\mathcal{N}_{zz}$ are instead unaffected by $U^{-1}$, and their behavior can be safely used to reveal the PS1
phase in experiments: the contrast of their modulations grows with $\Omega$ and reduces at increasing $|\varepsilon|$, vanishing asymptotically at large negative
values of the quadratic Zeeman field.

The energetic cost of density modulations increases with $\Omega$, until a transition occurs towards a plane-wave (PW) phase,
where all particles condense in a single minimum with momentum $k_1$. In this phase the density is uniform and the longitudinal
magnetization is fixed by the momentum, as $\langle F_z \rangle = k_1/k_0 > 0$. The degeneracy with the state having opposite
momentum (and magnetization along $z$) persists.

In a very narrow region between the ZM phase (above) and the PS1 and PW phases (below), a second polar striped (PS2) phase
appears, featuring occupation of both odd and even-$l$ momentum states in~\eqref{eq:ansatz}. In the PS2 phase, having $\langle F_z \rangle=0$,
the densities $\mathcal{F}_z$, $\mathcal{N}_{zx}$, and $\mathcal{N}_{yz}$ oscillate with period $2\pi/k_1$, while $n$, $\mathcal{F}_x$,
$\mathcal{F}_y$, $\mathcal{N}_{xx}$, $\mathcal{N}_{yy}$, $\mathcal{N}_{xy}$, and $\mathcal{N}_{zz}$ have oscillation wavelength
$\pi/k_1$ as in the PS1 phase (see Supplemental Material~\cite{Supplemental}).

The PS1-PS2 and PS2-ZM phase transitions have a second-order nature, involving a smooth variation of the contrast of the stripes
and of the momentum distribution. The PS1-PW and PS2-PW transitions are instead of first order, and both the contrast
of the stripes and the magnetization $\langle F_z \rangle$ can be used as experimentally testable order parameters.

The PS1-PS2, PS1-PW, and PS2-PW transition lines meet at the tricritical point $C_P$, where the three phases coexist.
The appearance of $C_P$ is an effect of the antiferromagnetic spin-dependent interactions and represents a key feature predicted
by our work. Interactions also turn $C_S$, which in the noninteracting model connects three different single-particle regimes~\cite{Lan2014},
into a second tricriticality, where the PS2-ZM and PS2-PW lines intersect. After $C_S$, a direct PW-ZM transition occurs,
whose second-order nature is witnessed by a smooth behavior of the magnetization $\langle F_z \rangle$. Up to corrections due to the spin-dependent
interaction, the PW-ZM transition follows the line separating the nondegenerate ($p_x=0$) and twofold-degenerate ($p_x=\pm k_1$) regimes of the
single-particle ground state~\cite{Lan2014}.

For a fixed $\bar{n}$, as $g_2\to 0^{+}$ the point $C_P$ shifts toward the origin $(\Omega,\varepsilon)=(0,0)$, while $C_S$, whose position
generally depends on interactions, approaches its noninteracting counterpart. In the same limit, the region of the PS1 phase collapses
into the negative $\varepsilon$ half-axis, and the PW phase, which is uniform and favored by density-density interactions, extends in the whole
lower region of the phase diagram. The PS2 phase instead shrinks into the same line which, in the noninteracting limit, separates the nondegenerate
and twofold-degenerate single-particle regimes, and connects the origin with $C_S$~\cite{Lan2014}.

For a given value of $\varepsilon$  the critical Raman coupling $\Omega$ characterizing the PS1-PW transition approaches, as $\bar{n} \to 0$, 
a density-independent value, fixed by the ratio $g_2/g_0$, as in the spin-$1/2$ case~\cite{Ho2011}. Remarkably, also the tricritical point $C_P$
approaches a finite value as $\bar{n} \to 0$. The behavior of the PS2 phase in the low-density limit is the same as for $g_2\to 0^{+}$.
A different scenario takes place for large densities: as $\bar{n}$ increases, $C_P$ approaches $C_S$ while at the same time the PW phase shrinks
and eventually disappears from the phase diagram, again as in the spin-$1/2$ case~\cite{Li2012PRL}. However, the merging of the two tricritical
points requires exceedingly high densities.

\begin{figure}
\includegraphics{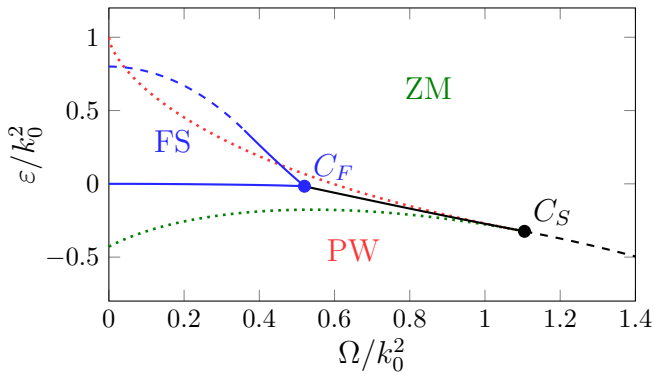}
\caption{Phase diagram for \emph{ferromagnetic interactions} for $g_0\bar{n}/k_0^2 = 0.4$ and $g_2\bar{n}/k_0^2 = - 0.2$.
The ratio $g_2/g_0 = -0.5$ is typical of ${}^7$Li. Continuous (dashed) lines represent first-(second-)order transitions.
The blue lines define the region of the ferromagnetic striped (FS) phase. The black line represents the transition between the
zero-momentum (ZM) and plane-wave (PW) phases. At the dotted green (red) line, the ZM (PW) state disappears as a local minimum
of the energy.}
\label{fig:ferr_phase_diag}
\end{figure}

\textit{Ferromagnetic interactions ---} In the case of ferromagnetic spin-dependent interactions ($g_2<0$), three phases are possible
without SO coupling~\cite{Ho1998,Ohmi1998,StamperKurn2013}. For $\varepsilon \geq 4|g_2|\bar{n}$, the ground state
is again the longitudinal polar state discussed above, while at $\varepsilon < 0$ the two degenerate longitudinal ferromagnetic states
$\Psi = \sqrt{\bar{n}} \, (1 , 0 , 0)^T$ and $\Psi = \sqrt{\bar{n}} \, (0 , 0 , 1)^T$ are energetically favored. In the intermediate regime
$0 < \varepsilon \leq 4|g_2|\bar{n}$, the ground-state wave function reads $\Psi = \sqrt{\bar{n}/2} \, (\sin\beta \, \me^{-\mi\chi} ,
\sqrt{2}\cos\beta , \sin\beta \, \me^{\mi\chi})^T$, with $\cos 2\beta = \varepsilon/|4g_2\bar{n}|$. This state is characterized by
ferromagnetic order in the transverse $x$-$y$ plane, with the magnetization direction fixed by the arbitrary phase factor $\chi$, and
continuously approaches the polar state at $\varepsilon=4|g_2|\bar{n}$, in a second-order phase transition. The transverse and longitudinal
ferromagnetic states are instead separated by a first-order transition.

The phase diagram in presence of SO coupling is shown in Fig.~\ref{fig:ferr_phase_diag}. At low $\Omega$, the ZM and PW phases
are still present in the upper and the lower part of the diagram, respectively. Their properties are essentially the same as in the
antiferromagnetic case; however, for $g_2<0$ the ground state is in the PW phase also at small Raman couplings, and approaches
the above longitudinal ferromagnetic phase as $\Omega \to 0$.

At low $\Omega$ and $\varepsilon$ another class of ground states is found, stemming from the transverse
ferromagnetic state at $\Omega=0$. In this ferromagnetic striped (FS) phase, the harmonic terms with $l=0,\pm 1$
in~\eqref{eq:ansatz} are predominantly populated, with $|C_0| > |C_{l \neq 0}|$. The FS phase results from
the competition between the spin-dependent interaction term, proportional to $g_2$, and the quadratic Zeeman field with
strength $\varepsilon$. It has $\langle F_z\rangle = 0$ and, due to its connection with the transverse ferromagnetic phase
at $\Omega = 0$, the contrast of the density modulations is larger for the magnetization densities $\mathcal{F}_x$ and
$\mathcal{F}_y$ than for the total density $n$ and the nematicity densities $\mathcal{N}_{xx}$, $\mathcal{N}_{yy}$, $\mathcal{N}_{xy}$,
and $\mathcal{N}_{zz}$. All these quantities oscillate with wavelength $2\pi/k_1$, twice larger than in the PS1 phase for $g_2>0$~\cite{Lan2014},
with relevant anharmonicities for the nematicity densities (see Supplemental Material~\cite{Supplemental}).
As for the PS1 phase, in the FS phase the maximum contrast in $n$ at fixed $\Omega$ is achieved by decreasing $|\varepsilon|$,
i.e., going away from the upper transition line. However, the behavior of the contrast versus $\Omega$ is not monotonic. Concerning
the nematicity density $\mathcal{N}_{zz}$, its contrast is generally neither monotonic in $\Omega$ nor in $\varepsilon$.

The PW-FS transition has a first-order nature, which can be revealed by looking at the jumps in $\langle F_z \rangle$ and in
the contrast of the density modulations of the FS phase. The behavior of the contrast also witnesses the order of the FS-ZM
transition. At small values of $\Omega$, the transition is expected to be second order as for $\Omega=0$. However, for our
values of the parameters we find that, at larger $\Omega$, the contrast suddenly jumps from zero to a finite value, indicating
the presence of a first-order phase transition.

The PW-FS and FS-ZM transition lines meet at the tricritical point $C_F$, which exists as a consequence of the ferromagnetic
spin-dependent interactions, and whose distance from the origin increases with $|g_2|/g_0$. Passing $C_F$ one
has a direct PW-ZM transition, which is of first order at small Raman couplings. However, the jump in the order parameter
$\langle F_z\rangle$ decreases when moving towards higher $\Omega$, while the region of the $\Omega$-$\varepsilon$ plane
where both the PW and ZM phases are present as local minima of the energy shrinks (see Fig.~\ref{fig:ferr_phase_diag}).
Eventually, at the critical point $C_S$, the PW-ZM transition becomes second order. Notice the difference in the nature of $C_S$
with respect to both the antiferromagnetic case and the noninteracting model~\cite{Lan2014}. The change in the order
of the PW-ZM transition has already been confirmed in experiments with ferromagnetic ${}^{87}$Rb BECs~\cite{Campbell2016}.

For a fixed ratio $g_2/g_0$, as the average density $\bar{n}$ increases, the region of the FS phase in the phase diagram
grows indefinitely along $\varepsilon$, and the position of $C_F$ shifts to the right. In the large-$\bar{n}$ regime $C_F$ can meet
and cross $C_S$, which moves leftward, and the whole PW-ZM transition becomes second order. In the opposite $\bar{n} \to 0$
limit, the FS phase shrinks into a finite arc of the curve connecting the origin with $C_S$, similar to the PS2 phase in the antiferromagnetic
case. At fixed $\bar{n}$, and for $g_2 \to 0^{-}$, the region of the FS phase collapses, together with $C_F$, into the origin of the
$\Omega$-$\varepsilon$ plane.

\textit{Conclusions ---} Let us now comment on the experimental relevance of our results. The phase diagram for antiferromagnetic
interactions could be explored in experiments with ${}^{23}$Na spin-$1$ BECs. The transition from the ZM to the striped phases
can be studied by observing the change in the momentum distribution of the condensate, while the PW-ZM, the PS1-PW, and the PS2-PW
transitions can be revealed measuring the longitudinal magnetization $\langle F_z\rangle$. The existence of the tricriticalities
$C_P$ and $C_S$ can be deduced from the simultaneous study of these phase transitions. The possibility of tuning $\Omega$ and
$\varepsilon$ arbitrarily close to such points opens realistic perspectives for the first experimental detection of tricriticality
in ultracold atomic gases.

Similar considerations as above hold for ferromagnetic interactions. However, in current experiments with ${}^{87}$Rb BECs
the investigation of the FS phase is made difficult by the small ratio $g_2/g_0\simeq -0.005$~\cite{StamperKurn2013}, which compresses
the stripes in a narrow region of the $\Omega$-$\varepsilon$ plane. Possible strategies to improve the situation could involve atomic
species with larger or tunable values of $g_2/g_0$, or the implementation of properly chosen spin-dependent trapping
conditions~\cite{Martone2014}; this would also make the merging of $C_P$ or $C_F$ with $C_S$ of more realistic achievement.

Further developments concern the study of the dynamic properties of spin-$1$ configurations, including the behavior of the roton minima
in the excitation spectrum~\cite{Martone2012,Zheng2013,Khamehchi2014,Ji2015} and the emergence of a double gapless band structure
in the striped phases~\cite{Li2013}.

\begin{acknowledgments}
Insightful discussions with Giacomo Lamporesi, Yun Li, Chunlei Qu, and Zeng-Qiang Yu are acknowledged.
This work was partially supported by PRIN Grant No.~2010LLKJBX ``Collective quantum phenomena: from strongly correlated systems
to quantum simulators,'' by INFN through the project ``QUANTUM,'' by the Italian National Group of Mathematical Physics (GNFM-INdAM),
by the QUIC grant of the Horizon2020 FET program, and by Provincia Autonoma di Trento.

\textit{Note added ---} During the final preparation of the present work we became aware of two papers~\cite{Sun2016,Yu2016}
discussing spin-$1$ SO-coupled BECs, whose predictions, when comparable, are in agreement with our findings.
\end{acknowledgments}

\end{document}